\documentstyle[12pt]{article}
\newcommand{\gsim}{ \raisebox{-.5ex}{\mbox{$\,\stackrel{>}{\sim}\,$}} }
\newcommand{\lsim}{ \raisebox{-.5ex}{\mbox{$\,\stackrel{<}{\sim}$\,}} }
\begin{document}
\begin{center}
{\large\bf PSEUDOCHAOS IN STATISTICAL PHYSICS}\footnote{A talk given at the
Intern. Conference "Nonlinear Dynamics, Chaotic and Complex Systems",
Zakopane, 1995}\\[2mm]

                   Boris Chirikov\\
{\it Budker Institute of Nuclear Physics \\
        630090 Novosibirsk, Russia \\
          chirikov @ inp.nsk.su}\\[5mm]
\end{center} 
\baselineskip=15pt

\vspace{2mm}

\begin{abstract}
A new generic dynamical phenomenon of {\it pseudochaos} and its relevance
to the statistical physics both modern as well as traditional one are
considered and explained in some detail. 
The pseudochaos is defined as a statistical behavior of the dynamical system
with {\it discrete} energy and/or frequency spectrum. In turn, the
statistical behavior is understood as time--reversible but nonrecurrent
relaxation to some steady state, at average, superimposed with irregular
fluctuations.
The main attention is payed to the most important and universal example
of pseudochaos, the so--called {\it quantum chaos} that is dynamical
chaos in bounded mesoscopic quantum systems. The quantum chaos as a
mechanism for implementation of the fundamental correspondence principle 
is also discussed.

The quantum relaxation localization, a peculiar characteristic implication
of pseudochaos, is reviewed in both time--dependent and conservative
systems with special emphasis on the {\it dynamical decoherence} of
quantum chaotic states. Recent results on the peculiar global structure 
of the energy shell, the Green function spectra and the eigenfunctions, 
both localized and ergodic, in a generic conservative quantum system 
are presented.

Examples of pseudochaos in classical systems are given including
linear oscillator and waves, digital computer and completely integrable 
systems. A far--reaching similarity between the dynamics of a few--freedom
quantum system at high energy levels ($n\to\infty$) and that of many--freedom
one ($N\to\infty$) is also discussed.
\end{abstract}

\newpage

\section{Introduction: the second birth of pseudochaos}

The conception of {\it pseudochaos} has been first explicitly introduced
[1] in attempts to interpret a very controversial phenomenon of
{\it quantum chaos}, and to understand its mechanism and physical meaning.
The term itself has been borrowed from the theory of the
well--known 'pseudorandom number generators' in a digital computer.
Even though such imitation of the 'true' random quantities is widely used
in many 'numerical experiments', e.g., ones employing the Monte--Carlo
techniques, this pseudorandomness was always considered as a very 
specific mathematical model of no general interest for the fundamental
physics. However, in recent numerous attempts to understand quantum chaos,
which is attracting ever growing attention of many researchers (see, e.g.,
Proceedings of International Conferences [2--4], and a Collection of
papers [5]), it is becoming more and more clear that this 'specific mechanism'
provides, in fact, a typical chaotic behavior in physical systems. 

Moreover, from the viewpoint of fundamental physics,  the pseudochaos is
the only kind of chaos principally possible in physical systems of finite
dimensions. In infinite macroscopic systems of traditional statistical 
mechanics (TSM), both {\it classical and quantal}, particularly,
in the principal TSM conception of the thermodinamic limit $N\to\infty$,
where $N$ is the number of freedoms, there is no such problem.
Namely, it has been rigorously proven (see, e.g., Ref.[6]) that, loosely
speaking, the 'true' chaos is a generic phenomenon in this limit
even if for any finite $N$ the system is completely integrable!

The discovery of dynamical chaos in finite (and even few--dimensional)
{\it classical} systems -- a fundamental breakthrough in recent decades --
has crucially changed the classical statistical mechanics. By now, this
new mechanism for the statistical laws is well understood ( but still not
very well known), and received the firm mathematical foundations in the
modern ergodic theory [6].

In all success of the latter a 'minor' problem still remains: such a mechanism 
does not work in finite {\it quantum} systems that is ones whose motion is 
bounded in the {\it phase space} and, hence, whose energy and frequency
spectrum is discrete. 

The simplest solution of this problem, which seems to be almost commonly
accepted currently, is that the dynamical chaos in such systems 
is simply impossible.
However, this seemingly obvious 'solution' is, in fact, a trap as it
immediately leads to a sharp and very profound contradiction with the 
fundamental correspondence principle [7]. We need to choose what to
sacrifice, this principle or the 'true' (=classical) chaos. I prefer 
the latter. If the phenomenon of quantum chaos did really violate
the correspondence principle as some physicists suspect
it were, indeed, a great discovery since it would mean that the
classical mechanics is not the limiting case of quantum mechanics but a 
different separate theory.  `Unfortunately', there
exists a less radical (but also interesting and important) resolution of
this difficulty -- the pseudochaos -- which is the main topic of my talk.

Within such a philosophical framework
the central physical problem is to understand
the nature and mechanism of dynamical chaos in quantum mechanics. 
In other words, we need the {\it quantum} theory of dynamical chaos
including the transition to the classical limit.
Certainly, the quantum chaos is a new dynamical phenomenon [7], related
but not identical to the classical chaos. 
 We call it {\it pseudochaos}, the term {\it pseudo}
intending to emphasize the difference from the 'classical' chaos
in the ergodic theory. From the physical point of view, I accept here,
the latter, strictly speaking, does not exist in the Nature. So, in the
common philosophy of the universal quantum mechanics the {\it pseudochaos
is the only true dynamical chaos}.
The classical chaos is but a limiting pattern which
is, nevertheless, very important both in the theory to compare 
with the real (quantum) chaos
and in applications as a very good approximation in macroscopic domain as is
the whole classical mechanics.
Ford calls it {\it mathematical chaos} as contrasted to the {\it real
physical chaos} in quantum mechanics [10]. 

I emphasize again that the classical chaos is impossible
in {\it finite and closed} quantum systems to which my talk is restricted.
Particularly, I am not going to discuss here the quantum measurement 
in which, by purpose,
the macroscopic (infinite--dimensional) processes are involved (see, e.g.,
Ref.[7]).

Thus, the physical meaning of the term 'pseudochaos' is principally
different (and even opposite) to that of 'pseudorandom numbers' in computer.
The reason for the latter, original, term 'pseudo' was twofold.
At the beginning, the first and apparently the only meaning was related
to the common belief that no dynamical, deterministic, system like computer
can produce anything random, by definition. This delusion has been overcome
in the theory of dynamical chaos on the field of {\it real} numbers.
However, the digital computer works on a finite lattice of {\it integers}.
This is qualitatively similar to a quantum behavior [11]. Computer numbers
as well as quantum variables can be, at most, 'pseudorandom' only 
as compared to the 'true' random classical quantities represented by
real numbers. But then, a very special notion of 'pseudorandom' is
sharply scrambling up to the level of a new fundamental conception 
in physics.

The quantum chaos is a part of quantum dynamics which, in turn, is a
particular class of dynamical systems. It became a hard physical problem
upon discovery and understanding of the classical dynamical chaos.
To explain the problem I need to briefly remind the main peculiarities
of the classical chaos, especially those crucial in the quantum theory.

\section{Asymptotic chaos in classical mechanics}

There are two equivalent description of classical mechanics or, more
generally, of any finite--dimensional dynamical systems:
via individual trajectories, and via a distribution function, or
phase--space density for Hamiltonian (most fundamental) systems. 

The trajectory obeys the motion equations, which are generally nonlinear,
and desribes a particular realization of system's dynamics in dependence
on the initial conditions. The phase density satisfies the Liouville
equation, which is always {\it linear} whatever the motion equations,
and which usually represents the typical (generic) dynamical behavior 
for a given system.
Particularly, all zero--measure sets of special trajectories are
automatically excluded. 

Notice, however, that in some special cases the phase density may display
the properties absent for trajectories. An interesting example [57] is
the correlation decay (and, hence, continuous spectrum) for a special
initial phase density in a completely integrable system. The point is
that such a decay is related to the correlation between {\it different}
trajectories rather than on the same ones. The trajectory spectrum remains
descrete, and the corresponding correlation persists. Particularly, this
explains a surprising phenomenon known as 'echo' which is the revival
of such correlations upon velocity reversal. It was observed in many
cases, and has several interesting applications (see, e.g., Ref.[63]).
An interesting open
question is the exact conditions for a phase density to represent the
{\it trajectory} properties which is the primary problem in dynamics.

The strongest statistical properties of a dynamical system are related
to the {\it local} exponential instability of trajectories, as described
by the {\it linearized} motion equations, provided the motion is {\it bounded}
in phase space. These two conditions are {\it sufficient} for a rapid
mixing of trajectories by the mechanism of 'stretching and folding'.
For the linear motion equations the combination of both is impossible
unless the whole phase space of the system is finite. A well--known example
of the latter is a model described by the {\it linear} 'Arnold cat map'[8]:
$$ 
   \begin{array}{ll}
   \overline{p}\,=\,p\,+\,x & mod\ 1 \\
   \overline{x}\,=\,x\,+\,\overline{p} & mod\ 1 
   \end{array} 
   \eqno (1)
$$
on a unit torus. The motion is exponentially unstable with (positive)
Lyapunov's exponent $\Lambda =\ln{[(3+\sqrt{5})/2)} > 0$, and is bounded by
operation $mod\ 1$. Notice that the linearized motion is described by
the same map but {\it without} $mod\ 1$ that is in {\it infinite} plane
($-\infty <dp,dx<\infty$). It is unbounded and {\it globally} unstable
but perfectly regular, the so--called hyperbolic motion:
$$
   dp\,=\,a\cdot\exp{(\Lambda t)}\,+\,b\cdot\exp{(-\Lambda t)}; \qquad
   dx\,=\,c\cdot\exp{(\Lambda t)}\,+\,d\cdot\exp{(-\Lambda t)}\,
   \eqno (2)
$$
where constants $a,b,c,d$ depend on the initial conditions and on $\Lambda$,
and where integer $t$ is discrete map's time. Remarkably, the motion (2)
is time--reversible but {\it unstable} in both senses ($t\to\pm\infty$).
This implies the time reversibility of all the statistical properties for
the main system (1), a surprising conclusion which is still confusing 
some researchers (see, e.g., Ref.[61]).

A nontrivial part of the relation between instability and chaos is in
that the instability must be {\it exponential}. A power--law instability
is insufficient for chaos. For example, if we change first Eq.(1) to
$\overline{p}=p$ the model becomes completely integrable with the
oscillation frequency depending on the motion integral $p$ (nonlinear 
oscillation). The latter produces {\it linear} (in time) instability but
the motion remains regular (with discrete spectrum). This is a typical
property of the completely integrable nonlinear oscillations [58] which
leads to a confusing difference in the dynamical behavior of trajectories
and of phase densities as mentioned above. Another open question is
how to choose the correct time variable for a particular dynamical problem
[7]. A change of time may convert the exponential instability into a
power--law one, and vice versa (see, e.g., Ref.[59] for discussion). 

The two above conditions for dynamical chaos can be realized in very simple
(particularly, few--dimensional) systems like model (1). Another simple
example, to which I will refer below, is the so--called 'kicked rotator'
described by the 'standard map' [1,7,9,11]:
$$
   \overline{p}\,=\,p\,+\,k\cdot\sin{x}; \qquad
   \overline{x}\,=\,x\,+\,T\cdot\overline{p} \eqno (3)
$$
also on a torus ($x,p\ mod\ 2\pi$) or on a cylinder ($x\ mod\ 2\pi\,,\ 
-\infty <p<\infty$). This model is well studied too, and has many physical
applications. The motion on cylinder is bounded in one variable only, yet
it is sufficient for chaos. 

The exponential instability implies continuous spectrum of the motion
which is equivalent, loosely speaking, to the mixing, or temporal 
{\it correlation
decay}. Apparently, this is the most important characteristic property
in the statistical mechanics underlying the principal and universal 
statistical phenomenon of {\it relaxation} to some steady state, or
statistical equilibrium. 

Aperiodic relaxation is especially clear in the Liouville picture for the
phase density behavior (see, e.g., Ref.[8]). Consider a certain basis for
Liouville's equation, for example
$$ 
   \varphi_{mn}\,=\,\exp{[2\pi\,i(mx\,+\,np)]} \eqno (4)
$$
where $m,n$ are any integers in a simple example of model (1). In other words,
we represent the phase density as a Fourier series:
$$
   f(x,\,p,\,t)\,=\,\sum_{m,n}\,F_{mn}(t)\,\varphi_{mn}(x,\,p)\,=\,
   \sum_{m,n}\,F_{mn}(0)\,\exp{[2\pi\,i(m(t)x\,+\,n(t)p)]}
   \eqno (5)
$$
Except $\varphi_{00}$ any other term of this series has zero total
probability, and characterizes the spatial correlation in the phase
density. Map (1) induces a map for the Fourier amplitudes, and for harmonic 
numbers:
$$
   \overline{F}_{mn}\,=\,F_{\overline{m}\,\overline{n}};    \qquad   
   \begin{array}{ll}
   \overline{n}\,=\,n\,+\,m \\
   \overline{m}\,=\,m\,+\,\overline{n} 
   \end{array} 
   \eqno (6)
$$
Remarkably, variables $m(t),\,n(t)$ obey the same map as for
the {\it linearized} motion equations in variables $dx,\,dp$ and with the same
instability rate $\Lambda$ on the {\it infinite} lattice ($m,n$). 
The dynamics of the phase density in the Fourier representation, described
by the same Eq.(2) (upon substitution of $m,n$ for $dx,dp$), is also unbounded,
globally unstable, and regular. 
This is of no surprise as both representations describe the {\it local}
structure of motion. Dynamical chaos is a {\it global} phenomenon determined,
nevertheless, by the microdetails of the initial conditions due to the
exponential instability of motion [7,13].
Accordingly,
in the original phase space the temporal density fluctuations 
are chaotic as are almost all trajectories of map (1).

The only {\it stationary} mode $m=n=0$
with the full probability represents in this picture the statistical steady
state while all the others describe {\it nonstationary} fluctuations.
The latter are another characteristic property of statistical behavior.
These can be separated from the average statistical relaxation 
by the so--called
coarse--graining, or spatial averaging, which is a projection of the phase 
density on a finite (and arbitrarily fine) partition of the phase space. 
The kinetic (particularly,
diffusive) description of the statistical relaxation is restricted to such a
coarse--grained projection only while the fluctuations work as a dynamical
{\it generator of noise}.

Another elegant method of separating out the average relaxation is the
suppression of the fluctuations using Prigogine's $\Lambda$ operator [61]
which provides an invertible smoothing of the exact phase
density [60]. True, the inverse operator in a nonproper one, yet this method
could be efficiently used in some theoretical constructions. 
Contrary to a common belief,
it has nothing to do with the time irreversibility [61,62]. Moreover,
unlike the coarse--grained projection the $\Lambda$--smoothed phase density
is as reversible as the exact one (principally but not practically, of
course). The origin of misunderstanding concerning 'irreversibility'
is apparently related to the necessary restriction on the initial smoothed
density which was missed in the theory [62]. Such a density is a technical,
rather than natural, property of the system, and hence it does not need
to be arbitrary.
 A similar operation is often used in quantum mechanics (for
different purposes) to convert the Wigner function (the counterpart of exact
classical phase density) into the so--called Husimi distribution which is
the expansion in the coherent states (see, e.g., Ref.[7]). 

Nonstationary fluctuations/correlations of the phase density
form a {\it stationary} flow into higher modes $|m|,\,|n|\to\infty$ 
(cf. Ref.[12]), and keep the memory of the exact initial conditions
(see first Eq.(6)) providing time reversibility for the exact density.
The stationary correlation flow is only possible for the 
{\it continuous} phase space 
which is a characteristic feature of classical mechanics. This
allows for the asymptotic ($t\to\pm\infty$) formulation of the ergodic theory.
 Notice that both trajectories and
the full density are {\it time--reversible} but the latter, 
unlike the former, is
{\it nonrecurrent}. Reversed relaxation, particularly 'antidiffusion',
describes the growth of a big fluctuation which is eventually 
(as $t\to -\infty$) followed
by the standard relaxation in the opposite direction of time [13].

\section{Quantum pseudochaos: a new dimension in the ergodic theory}

 The dynamical chaos is one limiting case of the modern general  
 theory of dynamical systems which describes  
 the statistical properties of deterministic motion (see, e.g., Ref.[6]).  
  No doubt, this  
 theory has been developed on the basis of classical mechanics. Yet, as a  
 general mathematical theory, it does not need to be restricted to classical 
 mechanics only. Particularly, it can be, and indeed was
 applied to quantum dynamics with a surprising result. Namely, as had  
 been found from the beginning[14] and was subsequently well confirmed (see,  
 e.g., Ref.[1,7,11,22,28]) the quantum mechanics does not typically permit 
 the 'true' (classical--like) chaos. This is because  
 in quantum mechanics the energy (and frequency) spectrum  
 of any system, whose motion is bounded in phase space, is discrete and  
 its motion is almost periodic.  
 Hence, according to the existing ergodic theory, such a quantum dynamics   
 belongs to the limiting case of regular motion which is opposite to 
 dynamical chaos. The ultimate origin of quantum  
 almost--periodicity is in the discreteness of the phase space itself  
 (or, in a more formal language, in the noncommutative geometry of the latter)
 which is at the basis of quantum physics and directly related  
 to the fundamental uncertainty principle.  
 Yet, another fundamental principle, the correspondence principle,   
 requires the transition  
 to classical mechanics in all cases including the  
 dynamical chaos with all its peculiar properties.  

Now, the principal question to be answered reads: where is the expected quantum
chaos in the ergodic theory? The answer to this question [7,11,13] 
(not commonly accepted as yet) was concluded from a simple observation 
(principally well--known but never comprehended enough) that the 
sharp border between the
discrete and continuous spectrum is
physically meaningful in the limit $|t|\to\infty$ only, the condition
actually assumed in the ergodic theory. Hence, to understand the quantum chaos
the existing ergodic theory needs some modification by introducing a new
'dimension', the time. In other words, a new and central problem in the ergodic
theory becomes the {\it finite--time statistical properties} of a dynamical 
system, both quantal as well as classical.

Within a finite time the discrete spectrum is dynamically equivalent to the
continuous one, thus providing much stronger statistical properties of the
motion than it was (and still is) expected in the ergodic theory in case of
discrete spectrum. It turns out that the motion with discrete spectrum 
may exhibit
{\it all} the statistical properties of the classical chaos but only on
some {\it finite} time scales. 
    
 The absence of the classical--like chaos in quantum dynamics apparently  
 contradicts not only the correspondence principle   
 but also the fundamental statistical nature of quantum mechanics. 
 However, even though the random element in quantum mechanics (`quantum
 jumps') is unavoidable, indeed, it can be singled out and separated from
 the proper quantum processes. Namely, the fundamental randomness in quantum
 mechanics is related only to a very specific event -- the {\it quantum
 measurement} -- which, in a sense, is foreign to the proper quantum system
 itself.  
 This allows to divide the whole problem of quantum dynamics in two
 qualitatively different parts:  

\begin{itemize}

 \item{} The proper quantum dynamics as described by a very specific   
 dynamical variable, the wavefunction $\psi (t)$ obeying some  
 deterministic equation, for example, the Schr\"o\-din\-ger equation.
The discussion below will be limited to this part only.

 \item{} The quantum measurement including the registration of   
 the result and, hence, the collapse of the $\psi$ function which
 still remains a very vague issue to the extent that there is no   
 common agreement even on the question whether this is a real   
 physical problem or an ill--posed one, so
 that the Copenhagen interpretation of quantum mechanics
 answers all the 'admissible' questions. In any event, there exists as yet no
 dynamical description of the quantum measurement including the $\psi $
 collapse.   

\end{itemize}

 Recent breakthrough in the understanding of quantum chaos has been
 achieved, particularly, due to the above philosophy of separating out
 the dynamical part of quantum mechanics. Such a philosophy is accepted,
 explicitly or more often implicitly, by most researchers in this field.   

\subsection{Time scales of pseudochaos}

The existing ergodic theory is asymptotic in time, and thus has no explicit 
time scales at all \footnote{Asymptotic statements in the ergodic theory
should not be always understood literally to avoid physical misconceptions
(see, e.g., Addendum in second Ref.[12]). Actually, the classical chaos
has also time scales, for example, a dynamical one ($\sim\Lambda^{-1}$) [7].}.
There are two reasons for this. One is technical: it is much 
simpler to derive the asymptotic relations than to obtain rigorous 
finite--time estimates. Another reason is more profound. All statements in
the ergodic theory hold true up to measure zero that is excluding some peculiar
nongeneric sets of zero measure. Even this minimal imperfection of the theory
did not seem completely satisfactory but has been 'swallowed' eventually and
is now commonly tolerated even among mathematicians, let alone
physicists. In a finite--time theory all these exceptions acquire a {\it small
but finite} measure which would be apparently 'unbearable' 
(for mathematicians).
Yet, there is a standard mathematical 'trick' for
avoiding both these difficulties. 

The most important time scale $t_R$ in quantum chaos is given by the general
estimate [7,11]:
$$ \ln{(\omega t_R)}\,\sim\,\ln{Q}\,, \qquad t_R\,\sim\,\frac{Q^{\alpha}}
   {\omega}\,\sim\,\rho_0\,\leq\,\rho_H \eqno (7)
$$
where $\omega$ and $\alpha\sim 1$ are system--dependent parameters,
and $Q\gg 1$ stands for some big (in semiclassical region) quantum parameter.
It may be, e.g., a quantum number $Q=I/\hbar$
related to a characteristic action variable $I$
or the total number of states for the bounded quantum motion in a phase
space domain of volume $\Gamma$:
$Q\approx\Gamma /(2\pi )^N$.

Here and below I set $\hbar =1$.

 This scale is called the
{\it relaxation time scale} referring to one of the principal properties of
the chaos -- {\it statistical relaxation} to some steady state.
The physical meaning of this scale is principally simple, and
is directly related to the fundamental uncertainty principle ($\Delta t
\cdot\Delta E\,\sim\,1$) as implemented in the second Eq.(7) where $\rho_H$
is the {\it full} average energy level density (also called Heisenberg time).
For $t\lsim t_R$ the discrete spectrum is not resolved, and the statistical
relaxation follows the classical (limiting) behavior. This is just 
the 'gap' in the
ergodic theory (supplemented with the additional dimension, the time) where
the pseudochaos, particularly quantum chaos, dwells. A more accurate estimate
relates $t_R$ to a {\it part} $\rho_0$ of the level density. This is the
density of the so--called {\it operative eigenstates} only, that is those
which are actually present in a particular quantum state $\psi$, and which do
actually control its dynamics. 

The formal trick mentioned above is to consider not finite--time relations, we 
really need in physics, but rather the special {\it conditional limit}:
$$ t,\,Q\,\to\,\infty\,, \qquad \tau_R\,=\,\frac{t}{t_R(Q)}\,=\,const 
   \eqno (8)
$$
where $\tau_R$ is a new dimensionless time.
The {\it double} limit (8) (unlike the single
one $Q\to\infty$) is {\it not} the classical mechanics which holds true, in
this representation, for $\tau_R\lsim 1$ and with respect to the statistical
relaxation only. For $\tau_R\gsim 1$ the behavior becomes essentially quantum
(even in the limit $Q\to\infty$ !) and is called nowadays {\it mesoscopics}.
Particularly, the quantum steady state is generally quite different from
the classical statistical equilibrium in that the former may be {\it localized}
(under certain conditions) that is {\it nonergodic} in spite of classical
ergodicity. 

Another important difference is in {\it fluctuations} which are
also a characteristic property of chaotic behavior. In comparison with 
classical mechanics the quantum $\psi (t)$ plays, in this respect, 
an intermediate
role between the classical trajectory with big relative
fluctuations $\sim 1$ and the coarse--grained classical phase density
with no fluctuations at all. Unlike both the fluctuations of $\psi (t)$,
or rather those of averages in a quantum state $\psi (t)$, 
are typically $\sim d_H^{-1/2}$ where 
$d_H$ is the number of operative eigenstates associated with quantum state
$\psi$ which the former may be also called the {\it Hilbert dimension} 
of state $\psi$.
In other words, chaotic $\psi (t)$ represents statistically
a {\it finite ensemble} of $\sim d_H$ independent systems even though formally
$\psi (t)$ describes a {\it single} system. 
The fluctuations clearly demonstrate 
the difference between physical time $t$ and auxiliary variable $\tau$:
in the double limit ($t,\,Q\to\infty$) the fluctuations vanish, and one needs
a new 'trick' to recover them for a finite $Q$.

The relaxation time scale should not be confused with the {\it Poincare
recurrence time} $t_P\gg t_R$ which is typically much longer, and which
sharply increases with decreasing the recurrence domain. Time scale
$t_P$ characterizes big fluctuations (for both the classical trajectory,
but not the phase density, and for the quantum $\psi$) 
of which recurrences
is a particular case. Unlike this, $t_R$ characterizes the average relaxation
process. Rare recurrences, the more rare the larger quantum parameter $Q$,
make quantum relaxation similar to the classical nonrecurrent one.

More strong statistical properties than relaxation and fluctuations are
related in the ergodic theory to the exponential instability of motion. The
importance of those stronger properties for the statistical mechanics 
is not completely clear [56]. Nevertheless,
in accordance with the correspondence principle, those stronger properties
are also present in quantum chaos as well but on a {\it much shorter} time
scale $t_r$:
$$ 
   \Lambda t_r\,\sim\,\ln{Q} \eqno (9)
$$
where $\Lambda$ is the classical Lyapunov exponent. This time scale was 
discovered and partly explained in Ref.[15] (see also Ref.[7,11]). 
We call it {\it random time scale}.
Indeed, according to the Ehrenfest theorem the motion of a narrow wave packet
follows the beam of classical trajectories as long as the packet remains
narrow, and hence it is as random as in the classical limit. 
Even though the random time scale is very short, it grows indefinitely as 
$Q\to\infty$. Thus, a
temporary, finite--time quantum pseudochaos turns into the classical dynamical 
chaos in accordance with the correspondence principle. 

Again, we may consider the {\it conditional limit}:
$$ t,\,Q\,\to\,\infty\,, \qquad \tau_r\,=\,\frac{t}{t_r(Q)}\,=\,
   const  \eqno (10)
$$
Notice that new scaled time $\tau_r$ is different from 
the previous one $\tau_R$ in Eq.(8).

Particularly, if we fix time $t$, then in the limit $Q\to\infty$ we obtain 
the transition to the classical instability in accordance with 
the correspondence 
principle while for $Q$ fixed, and $t\to\infty$ we have the proper quantum 
evolution in time. For example, the quantum Lyapunov exponent
$$ 
   \Lambda_q(\tau_r)\,\to\,\left\{\begin{array}{ll}
   \Lambda\, , \quad & \tau_r\,\ll\, 1 \\
   0\,,        \quad & \tau_r\,\gg\,1 \end{array} \right.  \eqno (11)
$$

The quantum instability ($\Lambda_q>0$) was observed in numerical experiments
[7,16]. What does terminate the instability for $t\gsim t_r$? 
A simple explanation is suggested by the classical picture of the phase
density evolution on the integer Fourier lattice $m,n$ discussed above
for model (1). Classical Fourier harmonics $m,n$ are of a kinematical nature
without any a priori dynamical restriction. Particularly, they can go, and
do so for a chaotic motion, arbitrarily large which corresponds to a
continuous classical phase space. On the contrary, the quantum phase space
is discrete.
At first glance, the quantum wave packet stretching/squeezing, similar
to the classical one, does not seem to be
principally restricted since only 2--dimensional area (per freedom) is
bounded in quantum mechanics. 
However, Fourier harmonics of the quantum phase density (Wigner function)
are directly related to the quantum dynamical variables, particularly,
to the action variables whose values are restricted by the quantum
parameter $Q$, hence estimate (9). In a simple model (1) this is related
to a finite size of the whole phase space. Generally, in a conservative
system with even infinite phase space the restriction is imposed by the energy
conservation. Numerical experiments reveal that the original 
wave packet, after a considerable stretching similar to the classical one,
is rapidly destroyed. Namely, it gets split into many new small packets [7,16].
The mechanism of this sharp 'disrupture' of the classical--like motion
is not completely clear (for a possible explanation see Ref.[7,17]). 
The resulting picture is qualitatively similar to that for the classical
phase density, the main difference being in the spatial fluctuation scale
bounded now from below by $1/Q$. Nevertheless, the quantum phase density
can be also decomposed into a coarse--grained average part, and the
fluctuations.
An important implication of this picture for the wave packet time evolution is
the rapid and complete destruction of the so--called generalized coherent
states [18] in quantum chaos.

In quasiclassical region ($Q\gg 1$) the scale $t_r\ll t_R$. This leads to a
surprising conclusion that the quantum diffusion and relaxation are
{\it dynamically stable} contrary to the classical behavior. It suggests, 
in turn, that the motion instability is, generally, not important {\it during}
statistical relaxation. However, the {\it foregoing} correlation decay on
the short time scale $t_r$ is crucial for the statistical properties
of quantum dynamics.

Dynamical stability of quantum diffusion has been proved
in striking numerical experiments with time
 reversal [19]. In a classical chaotic system the diffusion is immediately
 recovered due to numerical 'errors' (not random !) amplified by the local
 instability. On the contrary, the quantum 'antidiffusion' proceeds until the
 system passes, to a very high accuracy, the initial state, 
 and only then the normal
 diffusion is restored. The stability of quantun chaos on relaxation
 time scale is comprehensible as the random time scale is much shorter.
 Yet, the accuracy of the reversal (up to $\sim 10^{-15}$ (!) )
 is surprising. Apparently, this is explained
 by a relatively large size of the quantum wave packet as compared to the 
 unavoidable rounding-off errors unlike the classical computer trajectory
 which is just of that size [20]. In the standard map (3)
 (upon quantization) the size
of the optimal, least-spreading, wave packet $\Delta x\sim \sqrt{T}$ [11].
 On the other hand, any quantity in the computer must well exceed 
 the rounding--off error $\delta\ll 1$. Particularly, $T\gg\delta$, and 
 $(\Delta x)^2/\delta^2\gsim(T/\delta)\delta^{-1}\gg 1$.

\subsection{Classical--like relaxation and residual fluctuations}
            
The relaxation time scale $t_R$ is the most important of the two considered
above for two reasons. First, it is much longer than $t_r$, and second,
it is related to the principal process of statistical relaxation
which is the basis of statistical mechanics.
The short scale $t_r$ was interpreted in Ref.[15] (see also Ref.[26])
as a limit for the classical--like behavior of chaotic quantum motion.
Subsequently, it was found that the {\it method} of quasiclassical 
quantization can be extended on a much longer time [11,27]. However, the
{\it physics} on both time scales is qualitatively different: dynamical
instability on scale $t_r$, and statistical relaxation afterwards.

On the whole scale $t_R$ the discrete pseudochaos spectrum is not resolved,
and relaxation follows the classical law. Consider, for example, model (3),
the standard map on a torus with total number of quantum states $Q$,
and $p,x$ as the action--angle variables.

If perturbation parameter $k\gsim Q$ the relaxation to ergodic steady state 
in this model, as well as in model (1), is very quick, with characteristic
relaxation time $t_e\sim 1$ (iterations). Such regime does often take place
in many physical systems. Here I consider another, more interesting
for the problem of pseudochaos, case, namely the {\it diffusive relaxation} 
which occurs for a sufficiently weak perturbation
$$
   k\,\ll\,Q \eqno (12)
$$

In the classical limit this relaxation is descibed by the standard 
diffusion equation
$$
   \frac{\partial f(p,\,t)}{\partial t}\,=\,\frac{1}{2}\,
   \frac{\partial}{\partial p}\,D(p)\,\frac{\partial f(p,\,t)}{\partial p}   
   \eqno (13)
$$
where $f(p,t)=<f(p,x,t)>_x$ is a coarse--grained phase density (averaged
over $x$), and
$$
   D\,=\,\frac{<(\Delta p)^2>}{t}\,\approx\,\frac{k^2}{2} \eqno (14)
$$
the diffusion rate. The latter expression holds for the standard map if
 $K=kT\gg 1$ which is also the condition for the global chaos in 
 this model [9]. The relaxation to the ergodic steady state $f_s=1/Q$ 
 is exponential with characteristic time
$$
   t_e\,=\,\frac{Q^2}{2\pi^2\,D}\,\approx\,\frac{Q^2}{\pi^2\,k^2} 
   \eqno (15)
$$
In diffusive regime ($k\ll Q$) this time $t_e\gg 1$. That average relaxation is
stable and regular in spite of underlying chaotic dynamics. 

The quantized standard map $\overline{\psi}=\hat{U}\psi$ is described   
by a unitary operator
$$ \hat{U}\,=\,\exp{\left(\,-\,i\,\frac{T\,\hat{p}^2}{2}\right)}\,\cdot\,
   \exp{(\,-\,ik\cdot\cos{\hat{x}})} \eqno (16)
$$
on a cylinder ($Q\to\infty$) [14], where $\hat{p}=-i\,\partial /\partial x$, 
and by a similar but somewhat more complicated expression
on a torus [21].
 
There are three quantum parameters in this model: perturbation $k$, period $T$
and size $Q$ but only two classical combinations remain: 
perturbation $K=k\cdot T$, 
and classical size $M=TQ/2\pi$ which is the number of classical resonances 
over the
torus. Notice that the quantum dynamics is generally more rich than the
classical one as the former depends on an extra parameter. It is, of course,
another representation of Planck's constant which I have set $\hbar =1$.
This is why in the quantized standard map we need both parameters, $k$ and $T$,
separately and cannot combine them in a single classical parameter $K$.
 
The quasiclassical region, where we expect quantum chaos, corresponds to 
$T\to 0,\ k\to\infty ,\ Q\to\infty$ while the classical parameters 
$K=const$ and $M=const$ are fixed.

A technical difficulty in evaluating $t_R$ for a particular dynamical problem 
is in that the density $\rho_0$ depends, in turn, on the dynamics. So, we
have to solve a self--consistent problem. For the standard map the answer
is known (see Ref.[7]):
$$ t_R\,=\,\rho_0\,=\,2D \eqno (17)
$$
This is a remarkable relation as it connects essentially {\it quantum}
characteristics ($t_R,\ \rho_0$) with the {\it classical} diffusion rate
$D$ (14).

The quantum diffusion rate depends on the scaled (dimensionless) time
$\tau_R$ (8), and is given by 
$$
  D_q\,=\,\frac{D}{1\,+\,\tau_R}\,\to\,
  \left\{\begin{array}{ll}
  D\, , \quad & \tau_R\,=\,t/t_R\,\ll\, 1 \\
   0\,,        \quad & \tau_R\,\gg\,1 \end{array} \right.  \eqno (18)
$$
This is an example of scaling in discrete spectrum which 
eventually stops the quantum diffusion. 

The character of the steady state crucially depends on the ratio
$t_R/t_e$. Define the {\it ergodicity parameter} $\lambda$ as [7]
$$ \lambda\,=\,\frac{D}{Q}\,\sim\,\left(\frac{t_R}{t_e}\right)^{1/2}
   \,\sim\,\frac{k^2}{Q}\,\sim\,\frac{K}{M}\cdot k \eqno (19)
$$

Consider, first, the case $\lambda\gg 1$ when the time scale $t_R$ is long
enough to allow for the completion of the classical--like relaxation.
In this case the final steady state as well as all the eigenfunctions
are ergodic that is the corresponding Wigner functions
are close to the classical microcanonical distribution in phase space.
This region is inevitably reached if the classical parameter
$K/M$ is kept fixed while the quantum parameter
$k\to\infty$ in agreement with the Shnirelman theorem or, better to say, with a
physical generalization of that [23]. It is called the {\it far quasiclassical 
asymptotics}. 

The principal difference of the quantum ergodic state from the classical one
is {\it residual fluctuations} in the former.
In quasiclassical region the chaotic quantum steady state is a superposition
of very many eigenfunctions. As a result almost any physical quantity
fluctuates in time. Even in discrete spectrum we are considering here
these fluctuations are very irregular. 
In case of a classical--like ergodic steady state all $Q$ eigenfunctions
essentially contribute to the fluctuations. Moreover, we would expect
their contributions to be statistically almost independent. Hence, the
fluctuations should scale $\sim Q^{-1/2}=d_H^{-1/2}$ where $d_H$ is the 
Hilbert dimension of the ergodic state. This is the case, indeed, 
according to numerical experiments [29]. For example, the energy
fluctuations were found to follow a simple relation
$$ \frac{\Delta E_s}{E_s}\,\approx\,\frac{1}{\sqrt{Q}} \eqno (20)
$$
where
$$ (\Delta E_s)^2\,=\,\overline{E^2(t)}\,-\,E_s^2;
   \qquad   E_s\,=\,\overline{E(t)}; \qquad E(t)\,=\,\frac{<p^2>}{2}
   \eqno (21)
$$
Here the bar indicates time averaging over a sufficiently long time
interval ($\gg t_e$), and the brackets denote usual average
over the quantum state. 

Dependence (20) suggests the complete quantum decoherence in the final
steady state for any initial state even though the steady state is
formally a pure quantum state. For $Q\gg 1$ the fluctuations are small,
so that {\it statistically} the quantum relaxation is {\it nonrecurrent}.
The decoherence of a chaotic quantum state is also confirmed by the
independence (up to small fluctuations) of the final steady--state energy
$E_s$ on the initial $E(0)$. Since any particular initial quantum state 
is strongly coherent the decoherence is a result of the quantum chaos.
It is called {\it dynamical decoherence}. This is one of the most important 
results of the studies in quantum chaos.

\subsection{Mesoscopics: quantum behavior in quasiclassical region}

If ergodicity parameter $\lambda\ll 1$ is small all the eigenstates 
and the steady state are non--ergodic, or localized.
This is because the scale $t_R$ is not long enough to support the
classical--like diffusion which stops before the classical relaxation
is completed. For this reason it is also called the {\it quantum diffusion
localization}. As a result the structure of eigenfunctions and of the steady
state remains essentially quantum,
no matter how large is the quantum parameter $k\to\infty$. This is called
{\it intermediate quasiclassical asymptotics} or {\it mesoscopic} domain.
Particularly, it corresponds to $K>1$ fixed, $k\to\infty$ and 
$M\to\infty$ while $\lambda\ll 1$ remains small.

The popular term 'mesoscopic' means here some intermediate behavior
between classical and quantum one.
In other words, in mesoscopic phenomenona both classical and quantum
features are combined simulteneously. Again, the correspondence principle
requires transition to the completely classical behavior. This is, indeed, 
the case as the mesoscopic
phenomena occur in the region where quantum parameter $k\gg 1$ is already 
very big but still less than a certain
critical value (corresponding to $\lambda\sim 1$) which determines 
the border of transition to the fully classical
behavior (far quasiclassical asymptotics). 

If $\lambda\ll 1$ is very small
the shape of the localized eigenstates is asymptotically exponential [7],
and can be approximately described by a simple expression [41]:
$$
  f_m(p)\,=\,<|\psi_m(p)|^2>\,\approx\,\frac{2/\pi l}
  {\cosh{[2\,(p\,-\,p_m)\,/\,l]}} \eqno (22)
$$
The localized steady state has a similar but somewhat more complicated
shape [11,31]. This is a simple approximation
superimposed with big fluctuations.
 The parameter $l$ is called
{\it localization length}. Interestingly, the two localization lengths 
($l_s$ for steady state and $l$ for eigenfunctions) are rather
different [11]:
$$
  l_s\,\approx\,D \quad {\rm while} \quad l\,\approx\,\frac{D}{2}
  \eqno (23)
$$
because of big fluctuations.
  
In terms of localization length the region of mesoscopic phenomena
is defined by the double inequality:
$$ 1\,\ll\,l\,\ll\,Q \eqno (24)
$$
The left inequality is a classical feature of the state while
the right one refers to quantum effects.
The combination of both allows, particularly, for a classical description,
at least in the standard map, of statistical 
relaxation to the quantum steady state 
by a phenomenological diffusion equation [7,24]
for the Green function:
$$ \frac{\partial g(\nu,\,\sigma)}{\partial\sigma}\,=\,
   \frac{1}{4}\,
   \frac{\partial^2 g}{\partial\nu^2}\,+\,B(\nu)\,\frac{\partial g}
   {\partial\nu} \eqno (25)
$$
Here $g(\nu ,\,0)=|\psi (\nu ,0)|^2=\delta (\nu -\nu_0)$ and
$$ \nu\,=\,\frac{p}{2D}\,, \quad \sigma\,=\,\ln{(1\,+\,
   \tau_R)}\,,   \quad \tau_R\,=\,\frac{t}{2D} \eqno (26)
$$
The additional drift term in the diffusion equation with
$$ B(\nu )\,\approx\,{\rm sign}(\nu\,-\,\nu_0)\,=\,\pm 1 \eqno (27)
$$
describes the so--called quantum coherent backscattering which is the 
dynamical mechanism of localization.

The solution of Eq.(25) reads [7]:
$$
  g(\nu ,\,\sigma )\,=\,\frac{1}{\sqrt{\pi\sigma}}\,\exp{\left[\,-\,
  \frac{(\Delta\,+\,\sigma )^2}{\sigma}\right]}\,+\,\exp{(\,-\,4\Delta )}
  \cdot{\rm erfc}\left(\frac{\Delta\,-\,\sigma}{\sqrt{\sigma}}\right)
  \eqno (28)
$$
where $\Delta =|\nu -\nu_0|$.

Asymptotically, as $\sigma\to\infty$, the Green function $g(\nu ,\,\sigma )
\to\,2\,\exp{(-4\Delta )}\equiv g_s$ approaches the localized steady state
$g_s$, exponentially in $\sigma$ but only as a power--law in physical time
$\tau_R$ or $t$ ($g-g_s\sim 1/\tau_R$). This is the effect of discrete motion
spectrum. Numerical experiments confirm prediction (28), at least, to the
logarithmic accuracy $\sim\sigma\approx\ln{\tau_R}$ [7,22]. 

A physical example of localization is the quantum suppression of diffusive 
photoeffect in Hydrogen atom [25]. Depending on parameters the suppression
may occur no matter how large the atomic quantum numbers. This is a typical
mesoscopic phenomenon which had been predicted by the theory of quantum chaos,
and was subsequently observed in laboratory experiments.

One might expect that in case of localization ($D\ll Q$)
the fluctuations would scale like $l^{-1/2}\sim k^{-1}$ as the number of 
eigenfunctions coupled in the localized steady state is $\sim l$. 
This is, however,
{\it not} the case as was found already in first numerical experiments
(see Ref.[7]). According to more accurate data [29] the fluctuations are
described by the relation
$$ \frac{\Delta E_s}{E_s}\,=\,\frac{A}{k^{\gamma}}\,=\,
   \frac{a}{d_H^{\gamma /2}} \eqno (29)
$$
with fitting parameters $\gamma =0.55,\  a=0.65$. For a nonergodic state
the Hilbert dimension can be defined as (see Ref.[7])
$$
   d_H^{-1}\,=\,\frac{1}{3}\cdot\int\,|\psi (p)|^4\,\,dp\,=\,
   \int\,f^2(p)\,dp \eqno (30)
$$
where $f(p)$ is smoothed (coarse--grained) density, and the factor $1/3$
accouts for $\psi$ fluctuations [30]. In case of exponential localization (22) 
$d_H\approx \pi^2\,l/4$.
The most important parameter $\gamma$
here is about twice as small compared to the expected value $\gamma =1$. 
This result suggests some fractal properties of localized eigenfunctions
and/or of their spectra. To put it another way, a slow fluctuation decay (29)
implies {\it incomplete} quantum decoherence which can be characterized
by the number $d_s$ of statistically independent components in the steady
state [7]. Then, from Eqs.(20) and (29) we obtain in the two limits:
$$ 
  \frac{d_s}{d_H}\,\approx\,\left\{
  \begin{array}{ll} 1, \qquad  \qquad & \lambda\,\gg\,1 \\
                \frac{d_H^{\gamma -1}}{a^2}\,=\,2.4\,d_H^{-0.45}, \quad &
                \lambda\,\ll\,1 \end{array} \right.  \eqno (31)
$$
This result was confirmed in Ref.[31] for a band random matrix model.

The phenomenon of quantum diffusion localization explains also the 
limitation of quantum instability in systems with infinite phase space
like the standard map on a cylinder. Indeed, the maximal number of coupled 
states here is determined by the localization length whatever the total 
number of states in the system. Hence, we should substitute quantum
parameter $Q\sim l\sim k^2$ in estimate (9). Even if localization does not
take place (e.g., for standard map with parameter $k(p)$ depending on $p$,
see Ref.[7,11]),
so that the quantum diffusion doesn't stop at all and the quantum spectrum
becomes continuous, the number of coupled
states increases with time as a power law only ($\Delta p\sim \sqrt{t}$),
and hence the quantum Lyapunov exponent $\Lambda_q\to 0$ vanishes on the
relaxation time scale. Only if the action variables grow exponentially
the instability rate
$\Lambda_q$ remains finite, and the quantum chaos becomes asymptotical like
in the classical limit (see Ref.[11,32] for such 'exotic' models).

\subsection{Examples of pseudochaos in classical mechanics}

The pseudochaos is a new generic dynamical phenomenon missed in the ergodic
theory. No doubt, the most important particular case of pseudochaos is the
quantum chaos. Nevertheless, pseudochaos occurs in classical mechanics as well.
Here are a few examples of classical pseudochaos which may help to understand
the physical nature of quantum chaos.
Besides, it unveils new features of classical dynamics as well. 

{\bf Linear waves} is the most close to quantum mechanics example of
pseudochaos (see, e.g., Ref.[33]). I remind that here only a part of quantum 
dynamics is discussed, one described, e.g., by the Schr\"odinger equation
which is a linear wave equation. For this reason the quantum chaos is called
sometime wave chaos [34]. Classical electromagnetic waves are used in
laboratory experiments as a physical model for quantum chaos [35]. The
'classical' limit corresponds here to the geometrical optics, and the
'quantum' parameter $Q=L/\lambda$ is the ratio of a characteristic size $L$ 
of the system to wave length $\lambda$. As is well known in optics, no
matter how large is the ratio $\L/\lambda$ the diffraction patterns prevail at
a sufficiently far distance $R\gsim L^2/\lambda$. This is a sort of relaxation
scale: $R/\lambda\sim Q^2$. 

{\bf Linear oscillator} (many--dimensional) is also a particular 
representation of waves.
A broad class of quantum systems can be reduced to this
model [36]. Statistical properties of linear oscillator, particularly in the
thermodynamic limit ($N\to\infty$), were studied in Ref.[37] in the framework
of TSM. On the other hand, the theory of quantum chaos suggests more rich 
behavior for a big but finite $N$, particularly, the characteristic time
scales for the harmonic oscillations [38], the number of freedoms $N$
playing a role of the 'quantum' parameter.

{\bf Completely integrable nonlinear systems} also reveal pseudochaotic 
behavior. An example of statistical relaxation in the Toda lattice had been
presented in Ref.[39] much before the problem of quantum chaos arose. Moreover,
the strongest statistical properties in the limit $N\to\infty$, including one
equivalent to the exponential instability (the so--called $K$--property) were
rigorously proved just for the systems completely integrable
for any finite $N$ (see Ref.[6]). 

{\bf Digital computer} is a very specific classical dynamical system whose
properties are extremely important in view of the ever growing interest to
numerical experiments covering now all branches of science and beyond.
The computer is the 'overquantized' system in that {\it any} quantity here is
{\it discrete} while in quantum mechanics only the product of two conjugated
variables does so. 'Quantum' parameter here $Q=M$ which is the largest computer
integer, and the short time scale (9) $t_r\sim\ln{M}$, the number of
digits in the computer word [11]. Owing to the discreteness, any dynamical
trajectory in computer becomes eventually periodic, the effect well known 
in the theory and practice of pseudorandom number generators.
One should take all necessary precautions to exclude such computer artifacts
in numerical experiments (see, e.g., Ref.[40,64]). 
On the mathematical part, the periodic approximations
in dynamical systems are also considered in the ergodic theory, apparently
without any relation to pseudochaos in quantum mechanics or computer [6].

The computer pseudochaos seems to me most convincing argument for the 
researchers who are still reluctant to
accept the quantum chaos as, at least, a kind of chaos,
insisting that only the classical--like (asymptotic) chaos deserves
this name, the same chaos which was (and is) studied to a large extent just
on computer, that is the chaos inferred from a pseudochaos!

\section{Statistical theory of pseudochaos: \newline
         random matrices}

The complete solution of the dynamical quantum problem can be obtained via
diagonalization of the
Hamiltonian to find the energy (or quasi--energy) eigenvalues 
and eigenfunctions.
The evolution of any quantity is, then, expressed as a sum over the
eigenfunctions. For example, the energy time dependence is
$$
   E(t) = \sum_{mm^\prime}\,c_m \,c^*_{m^\prime}\, E_{mm^\prime}\, 
   \exp{[i\,(E_m\,-\,E_{m^\prime})\,t]} \eqno (32)
$$
where  $E_{mm^\prime}$ are the matrix elements, and the initial  state 
in momentum representation is
$\psi (n,\,0) = \sum\limits_m \, c_m\, \varphi_m(n)$. For chaotic motion 
the dependence is generally very complicated but 
the statistical properties of 
the motion can be inferred from the statistics of eigenfunctions $\varphi_m 
(n)$ (and hence of the 
matrix elements $E_{mm^\prime}$), and of eigenvalues $E_m$.

By now, there exists a well--developed random matrix theory (RMT, see, e.g.,
Ref.[43]) which describes the average properties of a
typical quantum system with a given symmetry of the Hamiltonian.
At the beginning the object of this theory was assumed to be a very 
complicated, particularly many--dimensional, quantum system 
as a representative of a certain statistical ensemble. With 
understanding the phenomenon of dynamical chaos it became clear that the 
number of system's 
freedoms is irrelevant. Instead, the number of quantum states (quantum
parameter $Q$) is 
of importance provided the dynamical chaos in the classical limit. 

This approach to the theory of complex quantum systems 
like atomic nuclei had been taken by Wigner [42] 
40 years ago, much before the problem of quantum chaos was realized.
He introduced the so--called band random matrices (BRM) which were most 
suitable to account for the structure of conservative systems. However,
due to severe mathematical difficulties, RMT
 immediately turned to a much simpler case of statistically homogeneous
(full) matrices, for which impressive theoretical results have been achieved
[43]. The price was that the full matrices describe the local chaotic
structure only, the limitation especially inacceptable in atoms [30,44].
Only recently the interest of some researchers turned back to the original
Wigner BRM [45,48].

One of the main results in the studies of quantum
chaos was the discovery of quantum diffusion localization as 
a mesoscopic quasiclassical
phenomenon. This phenomenon, discussed above, has been well studied and 
confirmed
by many researchers for the dynamical models described by maps. 
Contrary to a common belief the maps describe not only time--dependent
systems but also conservative ones in the form of 
Poincare maps (see, e.g., Ref.[49]). Nevertheless, 
to my knowledge no 
direct studies of the quantum localization in conservative systems
have been undertaken as yet,
either in laboratory or even in numerical experiments.
Moreover, the very existence of quantum localization in conservative
systems is challenged [50]. Here, I briefly describe
recent results concerning the structure of the
localized quantum chaos
in the momentum space of a generic few--freedom conservative system which
is classically strongly chaotic, particularly, ergodic on a compact energy
surface [41].

Generally, RMT is a statistical theory of the systems with discrete energy
spectrum. This is just the principal property of the quantum
dynamical chaos [7]. Thus, RMT turned out to become accidentally 
(!) a statistical 
theory for the coming quantum chaos. Remarkably,
this statistical theory does not include
any time--dependent noise that is any coupling to a thermal
bath, the standard element of most statistical theories.
Moreover, a {\it single matrix} from a given statistical ensemble
represents the typical (generic) {\it dynamical} system of a given class
characterized by the matrix parameters. This makes an important bridge
between dynamical and statistical description of the quantum chaos.
In matrix representation the similarity between the problem of quantum
diffusion localization in momentum space and the well--known dual
problem of Anderson localization in configurational space of 
disordered solids [53,54] is especially clear and instructive.

Consider real Hamiltonian matrices of a rather general type
$$
   H_{mn}\,=\,H_{nn}\,\delta_{mn}\,+\,v_{mn} \qquad m\,,n\,=\,1,...,N
\eqno (33)
$$
where off--diagonal matrix elements $v_{mn}=v_{nm}$ are random and 
statistically
independent with $<v_{mn}>=0$ and $<v_{mn}^2>=v^2$ for $|m-n|\leq b$, and are 
zero otherwise. The most important 
characteristic of these Wigner band random matrices (WBRM) is the average 
energy level density $\rho$ defined by the relation
$$
   \frac{1}{\rho}\,=\,\left<H_{mm}\,-\,H_{m'm'}\right> \eqno (34)
$$   
where $m'=m-1$.
The averaging here and below is understood either over disorder that is over
many random matrices or within a single sufficiently large matrix. Both
are equivalent owing to the assumed statistical independence 
of matrix elements.
In other words, many matrices are statistically equivalent to a big one.
Quantum numbers $m,\,n$ are generally arbitrary but we will have in mind
ones related to the action variables, thus considering the quantum
structure in the momentum space.
The basis in which the matrix elements are calculated is usually
assumed to correspond to a completely integrable system with $N$
quantum numbers where $N$ is the number of freedoms. By ordering
the basis states in energy
we can represent $N$ quantum numbers by
a single one related to the energy which is also an action variable.

In the classical limit
the definition of WBRM (33) corresponds to the standard Hamiltonian
$H\,=\,H_0\,+\,V$
where the perturbation $V$ is usually assumed to be sufficiently small
while the unperturbed Hamiltonian $H_0$ is completely integrable.

Quantum model (33) is defined by 3 independent physical 
parameters above:
$\rho ,\ v,$ and $b$. The fourth parameter, matrix size $Q$, is considered 
to be technical in this model provided $Q\gg d_e$ (see Eq.(38) below)
is big enough to avoid the boundary effects. 

In terms of unperturbed energy $E_0$ the classical chaotic trajectory 
of a given 
total energy $E=const$ fills up the {\it energy shell}
$ \Delta E_0 =\Delta V$ with the ergodic (microcanonical) measure $w_e$ 
depending on a particular perturbation function $V$. 
In the quantum system this measure characterizes the shape (distribution)
of the ergodic eigenfunction (EF) in the unperturbed basis. Conversly, 
if we keep fixed the 
unperturbed energy $E_0=const$, the measure $w_e$ describes the band of
energy surfaces $E=const$ whose trajectories reach the unperturbed energy 
$E_0$.
In a quantum system the measure $w_e$ in the latter case corresponds 
to the energy spectrum of a
Green function (GS) with initial energy $E_0$. This characteristic
was originally
introduced also by Wigner [42] as the 'strength function', the term
still in use in nuclear physics.
Now it is called also the 'local density of (eigen)states'.
   
For a typical perturbation,
represented by WBRM, $w_e$ depends on the Wigner parameter [42]
$q\,=\,(\rho \,v)^2/b$.
In the two limits [42] (see also Ref.[7,47])
$$
   w_e(E)\,=\,\left\{ 
   \begin{array}{ccc}
   \frac{2}{\pi E_{SC}^2}\,\sqrt{E_{SC}^2\,-\,E^2}, & |E|\,\le\,E_{SC}, 
   & q\,\gg\,1\\  & & \\
   \frac{\Gamma /2\pi}{E^2\,+\,\Gamma^2/4}\cdot
   \frac{\pi}{2\cdot\arctan{(1/\pi q)}}, & |E|\,\le\,E_{BW}, & q\,\ll\,1
   \end{array} \right. \eqno (35)
$$
provided $\eta=\rho v\gsim 1$ which is the condition for coupling 
neighboring  unperturbed states by the perturbation.
In opposite case $\eta\ll 1$ the impact of perturbation is negligible
which is called {\it perturbative localization}. The latter is a
well known {\it quantum} effect but not one we are interested in 
(for chaotic phenomena it was first considered in Ref.[51]).
What is less known that for the coupling of {\it all} unperturbed
states within the Hamiltonian band a stronger condition is required,
namely:
$$
   \eta\,\gsim\,\sqrt{b}\,, \quad {\rm or} \quad q\,\gsim\,1
   \eqno (36)
$$   
This is a simple estimate in the first order of perturbation theory.
Indeed, the coupling is $\sim V/\delta E$. Within the band
the energy detuning
$\delta E\sim b/\rho$ while the total random perturbation
$V\sim v\sqrt{b}$, hence estimate (36).
In opposite case $q\lsim 1$ a {\it partial perturbative localization}
takes place which is also a {\it quantum} phenomenon, and again not
one we have in mind speaking about the quantum localization.
The mechanism of perturbative localization is relatively simple
and straightforward. This quantum effect is completely absent only
in the first, semicircle (SC), limit of Eq.(35) where the 
width of the energy shell
$\Delta E = 2E_{SC}=2\sqrt{8\,b\,v^2}=4\sqrt{2q}E_b\gg E_b$, and
$E_b=b/\rho$ is the half--width of the Hamiltonian
matrix band in energy.
The latter inequality allows for {\it diffusive} quantum motion
within the energy shell as a single random jump is $\sim
b\ll\rho\Delta E$.
The quantum localization under consideration here is related just
to the localization (suppression) of quantum diffusion by the
interference effects in discrete spectrum (see, e.g., Ref.[7]).
Notice that the SC width immediately follows from the above estimate
for perturbative localization: $\delta E\sim\Delta E\sim v\sqrt{b}$.

In the second, Breit - Wigner (BW), limit of Eq.(35) the
full size of the energy shell $\Delta E = 2E_{BW}=2E_b$ is equal
to that of the Hamiltonian band. However, due to the partial perturbative
localization explained above the main peak of the quantum ergodic
measure is considerably more narrow, 
with the width $\Gamma = 2\pi\rho v^2 = 2\pi qE_b\ll E_b$. 
This is again in accordance with the same simple estimate:
$\delta E\sim\Gamma\sim v\sqrt{\rho\Gamma}$.

To the best of my knowledge, the quantum distributions (35) 
were theoretically derived and studied for GSa only.
Classically, the measure $w_e$ seems to be the same for both $E=const$ and
$E_0=const$ as determined by the same perturbation $V$.
One of the main recent results [41] in the studies of WBRM
is that the classical symmetry between EFs ($E=const$)
and GSa ($E_0=const$) is generally lost in quantum mechanics.
Namely, in ergodic case such a statistical symmetry still persists,
yet the quantum localization drastically violates the symmetry
producing a very intricate and unusual global structure of the
quantum chaos.

In a sense, the conservative system is always localized
(finite $\Delta E$) even for ergodic motion. This is the origin of 
misunderstanding sometimes (see, e.g., Ref.[52]). In fact, such
a {\it classical localization} is a trivial consequence
of energy conservation as was explained above. It persists, of course,
in the classical limit as well. Here we are interested in the {\it quantum
localization} explained above called simply localization below.

Similar to maps the localization in conservative systems also depends on the 
ergodicity parameter (cf. Eq.(19)):
$$
   \lambda\,=\,a\,\frac{b^2}{d_e}\,=\,\frac{ab^{3/2}}{4\sqrt{2}c\eta}
   \eqno (37)
$$   
Here the ergodicity corresponds not to the total number of states $Q$ as in
maps (19) but to that within the energy shell of width $\Delta E$:
$$
    d_e\,=\,c\cdot \rho\,(\Delta E)_{SC}\,=\,4c\eta\sqrt{2b} \eqno (38)
$$    
Hilbert dimension $d_e$ is also called the {\it ergodic localization 
length} as a measure of the maximal
number of basis states (BS) coupled by the perturbation in case of 
ergodic motion.
Numerical factor $c\approx 0.92$ is directly calculated from the limiting
expression (35) for a particular definition of $d$ in Eq.(30).
Relation (38) is valid formally in the SC region ($q\gg 1$) only but, 
according to
computations, has still the accuracy within a few per cent down
to $q\approx 0.4$.

Parameter $\lambda$ (37) had been found in Ref.[48] and 
implicitely used there (without any relation to ergodicity). It was
explained in details in Ref.[45] where factor $a\approx 1.2$
was also calculated numerically.

Localization is characterized by the parameter
$$
   \beta_d\,=\,\frac{d}{d_e}\,\approx\,1\,-\,{\rm e}^{-\,\lambda}\,<\,1
   \eqno (39)
$$   
Here $d$ stands for the actual average localization length of EFs
measured according to the same definition (30).
Empirical relation (39) has been found in numerical experiments [45]
to hold in the whole interval $\lambda\leq 2.5$ there, and was confirmed 
in Ref.[41] up to $\lambda\approx 7$.

In the BW region $ d_{e}=\pi\rho\Gamma =2\pi^2\,b\,q$, and
$\lambda\approx a\,b/(2\pi^2 q)\gg 1$ as $q\ll 1$ (35)
and $b\gg 1$ in quasiclassics.
Hence, localization is only possible
in the SC domain which was studied in Ref.[41].

The numerical results [41] were obtained from {\it two} individual
matrices: the main one for the localized case with parameters
$$ \lambda =0.23;\ q=90;\ Q=2400;\ v=0.1;\ b=10;\ \rho =300;\ 
   \eta =30;\ d_e=500
$$
and additional one for ergodic case with parameters
$$ \lambda =3.6;\ q=1;\ Q=2560;\ v=0.1;\ b=16;\ \rho =40;\ 
   \eta =4;\ d_e=84
$$

All results are entirely contained in the EF matrix 
$c_{mn}$ which relates the eigenfunctions $\psi_m$ to unperturbed 
basis states $\varphi_n$: 
$$
   \psi_m\,=\,\sum_n\,c_{mn}\cdot\varphi_n, \qquad
   w_m(n)\,=\,|\psi (n)|^2\,=\,c^2_{mn}\,=\,w_{mn} \eqno (40)
$$   
in momentum representation,
and in the corresponding string of eigenvalues $E_m\approx m/\rho$.
From the matrix $c_{mn}$ the statistics of both EFs as well as GSa
was evaluated. In order to suppress big fluctuations in individual 
distributions the averaging over $300$ of them
in the central part of the matrix was done in two different ways:
with respect to the energy shell center ('global average', localization
parameter (39) $\beta_d=\overline{\beta}_g$), and with
respect to the centers of individual distributions ('local average', 
$\beta_d=\overline{\beta}_l$). Besides, the average $<\beta_d>$ over 
$\beta_d$ values from the individual distributions was computed.

 In the ergodic case $\lambda =3.6$ both average
distributions for EFs are fairly close to the SC law - a remarkable result, 
because
 that law was theoretically predicted for the {\it other} distribution, 
that of GSa.
More precisely, the bulk ('cap') of the distributions are very
close to the limiting SC (35), except in the vicinity of 
the SC singularities.
Numerical values of the localization parameter ($\overline{\beta}_g=1.08,\ 
\overline{\beta}_l=0.94,\ <\beta_d>=0.99$) are 
in a reasonable agreement with scaled
$\beta_d=0.97$ for $\lambda =3.6$, Eq.(39).

As expected, the GS structure is similar:
$\overline{\beta}_g=1.07,\ \overline{\beta}_l=1.06,\ <\beta_d>=0.98$.

For finite $q$ all the distributions are bordered by the two
symmetric steep tails which apparently fall down even faster than the
simple exponential with a characteristic width $\sim b$. 
The physical mechanism of the tail formation 
is a specific quantum tunneling via intermediate BSs [30]. The asymptotic
theory of the tails was developed in Ref.[30,42,46]. Surprisingly, it works
reasonably well even near the SC borders.

The structure of matrix $c_{mn}$ is completely different in the 
localized case $\lambda =0.23$.
The EF {\it local} average shows clear evidence for exponential localization
with $\overline{\beta}_l=0.24$ which is again close to scaled
$\beta_d=0.21$ for $\lambda =0.23$.
However, the {\it global} average reveals a nice SC (with tails) in spite of 
localization ($\overline{\beta}_g=0.98$).
It shows that, at average, the EFs homogeneously fill up
the whole energy shell. In other words, their centers are randomly scattered
over the shell.

Unlike ergodic case the localized GS structure is quite different
from that of EFs. Both averages now yield
similar results which  well fit the SC distribution 
($\overline{\beta}_g=0.98,\ \overline{\beta}_l=0.96$ as compared with 
$\overline{\beta}_g=0.99,\ \overline{\beta}_l=0.24$ for EFs). So, GSa look
extended, yet they are localized! This is immediately clear from the third
average $<\beta_d>\,=0.20$.
The explanation of this apparent paradox is that even though the GSa are 
extended over the shell they are {\it sparse} that is
contain many 'holes'. 

In analysis of the WBRM structure theoretical expression (38) for the
ergodic localization length $d_e$ (the energy shell width) was used.
In more realistic and complicated physical models this might impede
the analysis. In this respect the new method for direct empirical
evaluation of $d_e$,
and hence the important localization parameters $\beta_d$ and $\lambda$
in Eq.(39), from both average distributions for GSa as well as from the
global average for EFs [41] looks very promising.

The physical interpretation of this structure based upon the
underlying chaotic dynamics is the
following. Spectral sparsity decreases the level density of the operative
EFs which is the main condition for quantum localization via decreasing
the relaxation time scale (see, e.g., [7]). Yet, the initial 
diffusion and relaxation
are still classical, similar to the ergodic case, which requires extended GSa.
On the other hand, EFs are directly
related to the steady--state density [11], both being solid because of a 
homogeneous diffusion during the statistical relaxation.

This picture allows to conjecture that for a classically {\it regular}
motion the EFs become also sparse, so that EF/GS symmetry is apparently
restored.

\section{Conclusion: pseudochaos and traditional \newline
         statistical mechanics}

The quantum chaos is a particular but most important example of the new
generic dynamical phenomenon -- {\it pseudochaos} in almost periodic motion. 
The statistical properties of the discrete--spectrum motion is not a 
completely new subject of research, it goes back to the time of intensive
studies in the mathematical foundations of statistical mechanics {\it before}
the dynamical chaos was discovered or, better to say, was understood (see, 
e.g., Ref.[55]). This early stage of the theory as well as the whole TSM
was equally applicable to both classical and quantum systems.
For the problem of pseudochaos one of the most
important rigorous results with far--reaching implications was the
{\it statistical independence} of oscillations with incommensurate 
(linearly independent) frequencies $\omega_n$, such that the only solution of
the resonance equation 
$$ \sum_n^N\,m_n\cdot\omega_n\,=\,0 \eqno (41)
$$
in integers is $m_n\equiv 0$ for all $n$. This is a generic property of the 
real numbers. In other words, the resonant frequencies (41) form a set 
of zero Lebesgue
measure. If we define now $y_n=\cos{(\omega_n t)}$ the statistical independence
of $y_n$ means that trajectory $y_n(t)$ is ergodic in $N$--cube $|y_n|\leq 1$.
This is a consequence of ergodicity of the phase trajectory $\phi_n (t)=
\omega_n t\ mod\ 2\pi $ on a torus $|\phi_n|\leq \pi$. 

Statistical independence is the basic property of a set to which 
the probability
theory is to be applied. Particularly, the sum of statistically independent
quantities
$$ x(t)\,=\,\sum_n^N\,A_n\cdot\cos{(\omega_n\,t\,+\,\phi_n)} \eqno (42)
$$
which is the motion with discrete spectrum, is a typical object 
of this theory.
However, the familiar statistical properties like Gaussian fluctuations,
postulated (directly or indirectly) in TSM, are reached in the thermodynamic 
limit $N\to\infty$ only [55]. In TSM this
limit corresponds to infinite--dimensional models [6] which provide a
very good approximation for macroscopic systems, both classical and quantal.

What is really necessary for good statistical properties of 
almost periodic motion (42)
is a big number of frequencies $N_{\omega}\to\infty$ which makes the discrete
spectrum continuous (in the limit). In TSM the latter condition is satisfied
by setting $N_{\omega}=N\to\infty$. The same holds true for quantum fields 
which are infinite--dimensional.
In the finite--dimensional quantum mechanics {\it another} mechanism, 
independent of $N$, works in the quasiclassical region $Q\gg 1$.
Indeed, if the quantum motion (42)
(with $\psi (t)$ instead of $x(t)$) is determined by many ($\sim Q$) 
eigenstates we can set $N_{\omega}=Q$ independent of $N$. The actual number of 
terms in expansion (42) depends, of course, on a particular state $\psi (t)$.
For example, if it is just an eigenstate the sum reduces
to a single term. This is reminiscent to some special peculiar trajectories of
classical chaotic motion whose total measure is {\it zero}. Similarly, in
quantum mechanics $N_{\omega}\sim Q$ for {\it most} states if the system is
{\it classically chaotic}.

For a regular motion in the classical limit the quantity $N_{\omega}
\ll Q$ becomes considerably smaller. For example (see Ref.[14]), 
in the standard map
$N_{\omega}=Q$ in ergodic case, $N_{\omega}\sim k^2$ in case of localization
(both classically chaotic, $K>1$) but only $N_{\omega}\sim k\ll k^2\lsim Q$
for classically regular motion ($K<1$). 
The quantum chaos--order transition
is not as sharp as the classical one but the ratio $N_{\omega}(K>1)/
N_{\omega}(K<1)\sim k\to\infty$ increases with quantum parameter $k$.

Thus, with respect to the mechanism of the quantum chaos we essentially
{\it come back} from the ergodic theory to old TSM with exchange 
of the number of freedoms $N$ for 
quantum parameter $Q$. However, in quantum mechanics we are not interested,
unlike TSM, in the limit $Q\to\infty$ which is simply the {\it classical} 
mechanics.
Here, the central problem is the statistical properties for {\it large but
finite} $Q$. This problem does not really exist in TSM describing 
macroscopic systems.
In a {\it finite--Q} (or finite--N) pseudochaos
we have to introduce the basic conception of {\it time scale} [11].
This allows for interpretation of quantum chaos as
a {\it new} dynamical phenomenon, related but not identical at all to the
classical dynamical chaos. Hence, the term {\it pseudochaos}
emphasizing the difference from the asymptotic (in time) chaos
in the ergodic theory. 
 
 In my opinion, the fundamental importance of quantum chaos is precisely   
 in that it reconciles the two apparently opposite regimes, regular and  
 chaotic, in the general theory of dynamical systems. 
 The studies in quantum chaos help to better understand the old mechanism for 
chaos in many--dimensional systems. Particularly, the existence
of characteristic time scales similar to those in quantum systems
was conjectured in Ref.[7].

Is pseudochaos a chaos?

Until recently even the conception of classical dynamical chaos was rather
incomprehensible, especially for physicists. 
I know that some researchers actually observed dynamical chaos
in numerical or laboratory experiments but... did their best to get rid of it
as some artifact, noise or other interference! Now the situation in this field
is upside down: many researchers insist that if an apparent chaos
is not like that in the classical mechanics (and in the existing 
ergodic theory),
then it is not a chaos at all. Hence, sharp
disputes over the quantum chaos. The peculiarity of the current 
situation is that
in most studies of the 'true' (classical) chaos the digital computer is used
where only {\it pseudochaos} is possible that is one like in {\it quantum}
(not classical) mechanics!

Hopefully, this 'child disease' of quantum chaos will be over before long...

\vspace{5mm}

{\bf Acknowledgment.} The conception of quantum chaos presented above
has been developed in a long--term collaboration with G. Casati, J. Ford,
I. Guarneri, F.M. Izrailev and D.L. Shepelyansky.

\end{document}